\documentclass[3p,times,twocolumn]{elsarticle}
\usepackage{amssymb,amsmath,multirow,dcolumn,bm,float,graphicx,epstopdf}
\usepackage{calrsfs}  
\usepackage{epsfig}
\usepackage{soul}
\makeatletter

\usepackage{color}
\usepackage{amssymb,bbold}
\usepackage{graphicx,color}
\usepackage[usenames,dvipsnames]{xcolor}
\usepackage[section]{placeins}
\usepackage{amsmath}
\usepackage{subfigure}
\usepackage[normalem]{ulem}
\usepackage{booktabs}
\usepackage{xcolor}
\usepackage{epsfig,graphics,color,graphicx,amsmath}
\colorlet{darkgreen}{green!50!black}
\colorlet{brightyellow}{yellow!75!red}
\colorlet{orange}{red!50!yellow}
\colorlet{darkblue}{blue!60!black}
\colorlet{darkred}{red!80!black}

\def\bwt{\begin{widetext}}
\def\ewt{\end{widetext}}

\usepackage{soul}

\usepackage{mathtools}
\usepackage{mathrsfs}
\usepackage{bm}
\usepackage{relsize}
\usepackage{indentfirst}

\usepackage{xcolor}

\usepackage{amssymb,amsmath,multirow,epsfig,graphicx,color}

\usepackage{epsfig}
\usepackage{graphicx,amsmath}
\def\be{\begin{eqnarray} &&}
\def\nonu{\nonumber \\ &&}

\def\ee{\end{eqnarray}}
\def\psla{\slash \! \!\! }

\begin{document}
\begin{frontmatter}

\title{Exploring the $0^-$ bound state with dressed quarks in Minkowski space }

\author[1]{A. Castro}
\ead{abigail@ita.br}
\address[1]{Instituto Tecnol\'ogico de Aeron\'autica,  DCTA,
12228-900 S\~ao Jos\'e dos Campos,~Brazil}
\author[1]{W. de Paula}
\ead{wayne@ita.br}
\author[2]{E.~Ydrefors}
\ead{ydrefors@kth.se}
\address[2]{Institute of Modern Physics, Chinese Academy of Sciences, Lanzhou 730000, China}
\author[1]{T.~Frederico}
\ead{tobias@ita.br}
\author[3]{G. Salm\`e}
\ead{Giovanni.Salme@roma1.infn.it}
\address[3]{INFN, Sezione di Roma, P.le A. Moro 2, 00185 Rome, Italy}

\date{\today}

\begin{abstract}
The  Bethe-Salpeter equation for a pseudoscalar bound-system, with i) a ladder kernel with massive gluons, ii)  dynamically-dressed quark  mass function and iii) an extended quark-gluon vertex, is solved in Minkowski space by using the Nakanishi integral representation of the Bethe-Salpeter amplitude. The quark dressing is implemented through a phenomenological ansatz,  which was tuned by lattice QCD calculations of the quark running mass.  The latter were also used for assigning the range of the gluon mass and the parameter featuring the extended color density. This framework allows to investigate the gluon dynamics that manifest itself in the quark dressing, quark-gluon vertex  and the binding,  directly in the physical space. We present the first results for low-density pseudoscalar systems in order to elucidate the onset of the interplay between the above mentioned gluonic phenomena, and we discuss both static and dynamical quantities, like valence longitudinal and transverse distributions. 
\end{abstract}

\end{frontmatter}

{\it Introduction}
Understanding the relation  between  dynamical chiral symmetry breaking (DCSB)~(see, e.g., Refs. \cite{Cloet:2013jya,Eichmann:2016yit,Roberts:2021nhw} for a general introduction)  and  bound-state formation in Minkowski space is a fundamental issue for shedding light on the generation of the hadronic masses. In fact, many valuable  theoretical and experimental efforts   are underway and/or planned in the near future (see, e.g., Refs. \cite{AbdulKhalek:2021gbh,Anderle:2021wcy} for the experimental state-of-the-art). The attempt to describe the hadronic bound-states in Minkowski space by using a dynamical equation,  living in a genuine relativistic quantum-field theory realm, deserves attention, in view of the possible cross-fertilization with the main tool represented by the lattice Quantum-Chromodynamics (LQCD) and also the continuum approach to QCD, extensively pursued in Euclidean space (see Ref. \cite{Roberts:2020udq,Qin:2020rad} for  recent overviews).
Our approach is based on the $q\bar q$ homogeneous Bethe-Salpeter  equation  (BSE) in Minkowski space  (see also Refs.~\cite{Sauli:2008bn,Biernat:2013fka,Lucha:2015hta}  for different approaches in Minkowski space), with phenomenologically dressed  quarks that  interact through a ladder exchange of  massive gluons. Therefore, we  do not rely on the equivalence between the relativistic quantum-field theory and the Euclidean version, well-established within the formal framework elaborated in the 70's by Osterwalder and Schrader (see Ref.~\cite{Osterwalder:1973dx,Osterwalder:1974tc} where   necessary and sufficient conditions  for ensuring the validity of the Wick rotation are demonstrated). In what follows,  we adopt the well-established Nakanishi integral representation (NIR) of the Bethe-Salpeter (BS) amplitude and the light-front (LF)  projection of the BSE, in order to obtain a fully equivalent system of integral equations that allows to determine the Nakanishi weight functions (NWFs), and eventually the BS amplitude (see  Refs.~\cite{dePaula:2016oct,dePaula:2017ikc} for the general application to a $0^-$ bound system with undressed propagator, and Refs.~\cite{dePaula:2020qna,Ydrefors:2021dwa,dePaula:2022pcb,Ydrefors:2023src} for the weak decay constant,  valence probability, Ioffe distribution, electromagnetic form factor, parton distribution function  and   unpolarized transverse-momentum-dependent distribution functions for a pion with mass $m_{\pi}$ equal to $140$ MeV). After assigning the mass of the bound system, one solves the BSE and gets the corresponding coupling constant and  BS amplitude, that in turn allows one to  calculate the valence longitudinal and transverse momentum distributions. Both static and dynamical quantities are used for investigating how the interplay between  three fundamental gluonic phenomena, i.e. quark dressing, extended quark-gluon vertex and binding, evolves in a system-mass range $3m_\pi<M<5m_\pi$, corresponding to a relatively diluted  bound system with constrained quark-masses (see below).
The primary focus is on the combined effect of the quark dressing and   the increase of the color-density  extension around the quark-gluon vertex, which allows to enhance gluon fluctuations with low-momentum.

{\it The homogeneous BSE for a $0^-$ system.}
In the adopted ladder approximation with a massive gluon-exchange, the BSE for a $0^{-}$ quark-antiquark bound state, with total momentum $P$ and mass $M$,  is given by ~\cite{Salpeter1951} (see also Ref. \cite{GellMann:1951rw})
\begin{multline}\label{eq:bse1}
\Phi(k,P)=S \left(k+{P\over 2} \right)\int\frac{d^4k'}{(2\pi)^4}~D^{\mu\nu}(q) \Gamma_{\mu}(q)\\ \times~\Phi(k',p)~\hat{\Gamma}_{\nu}(q)~ S\left(k - {P\over 2} \right) \, ,
\end{multline}
where $k=(p_q - p_{\bar{q}})/2$, with $p_{q(\bar{q})}$ the off-shell (anti-) quark momentum, and $q = k - k'$. $\Gamma_{\mu}$ is the quark-gluon vertex and ${\hat{\Gamma}}_{\mu} = C\, \Gamma^T_{\mu} \, C^{-1}$, where $C = i \, \gamma^2 \, \gamma^0$ is the charge-conjugation operator. In Eq.~\eqref{eq:bse1}, we inserted the following three relevant ingredients. First,  a dressed Dirac propagator, that  for a parity-conserving interaction, reads 
\begin{multline}
    S(p)= S^V(p^2) \psla p + S^S(p^2)
\\ =
    i\int_{0}^{\infty}ds~\frac{\rho^V(s)\, \psla p+\rho^S(s)}{p^2-s+i\epsilon}~,
    \label{eq:KLR}
\end{multline}
where in the first line one has the decomposition in terms of allowed Dirac structures and  corresponding scalar functions,  while in the second line one has the dispersive representation (see, e.g., Ref.~\cite{Itzykson1980} for the K\"allen-Lehman representation in  QED), with  two  weights  to be phenomenologically parametrized  (at the present stage), as explained below. Second,
 a massive gluon-propagator, $D^{\mu\nu}(q)$, in the Feynman gauge, written as follows
 \begin{equation}
     D^{\mu\nu}(q)=-i {g^{\mu\nu} \over (q^2-\mu^2+i\epsilon)} \, , 
    \label{eq:glprop}
\end{equation}
where $\mu$ is the gluon mass. Last but not least, a quark-gluon vertex, $\Gamma^\mu(q)$, dressed through a simple form
factor, is  adopted, i.e.   
\be
\Gamma^{\mu}(q) = i \, g\, \frac{\mu^2-\Lambda^2}{q^2-\Lambda^2+i\epsilon} \, \gamma^{\mu} \, ,
\ee 
where $\Lambda$ features the extension of the color distribution dressing the interaction vertex.

For a $0^{-}$ $q\bar q$-system, the  BS amplitude can be decomposed as follows \cite{Carbonell:2010zw}
\begin{equation}\label{eq:4.114a}
\Phi(k,P)=\sum^{4}_{i=1}S_i(k,P)\phi_i(k,P),
\end{equation}
where i) $\phi_i$ are scalar functions, that under a $k\to -k$ transformations are even for $i=1,2,4$ and odd for $i=3$ (assuming {isospin} symmetry), and  ii)  the $S_i$   matrices compose the following orthogonal basis
\be
S_1(k,P)=\gamma^5 \, , \quad 
S_2(k,P)=\frac{\psla{P}}{M}\gamma^5 \, ,
\nonu
S_3(k,P)=\frac{k\cdot{P}}{M^3}\psla{P}\gamma^5-\frac{\psla{k}}{M}\gamma_5
\, , \quad \nonu
 S_4(k,P)=\frac{i\sigma^{\mu\nu} P_{\mu}k_\nu}{M^2}\gamma^5. 
\label{diracbasis}
\ee

\begin{figure}[t]
\begin{center}
\includegraphics[width=6cm, angle=0]{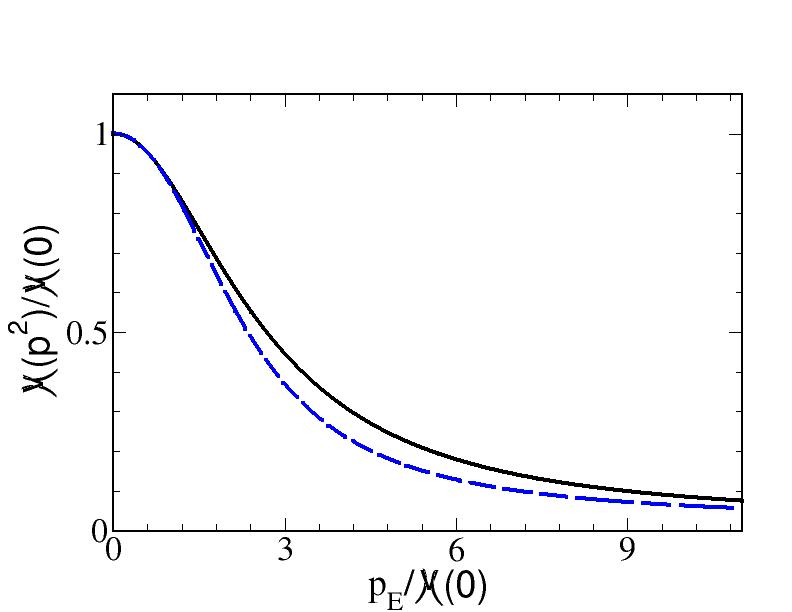}
\includegraphics[width=6cm, angle=0 ]{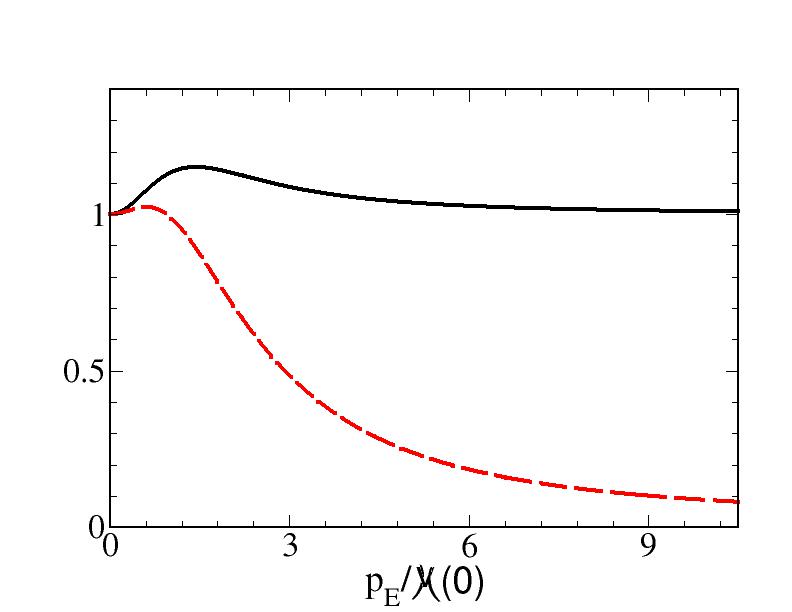}
\end{center}

\vspace{-0.5cm}

\caption{(Color online) Upper panel. The quark running-mass, ${\cal M}(p^2)$, as a function of the Euclidean momentum ${p_E=\sqrt{-p^2}}$, in units of the IR mass ${\cal M}(0) = 0.344\,$GeV.  Solid line: Eq.~\eqref{eq:Mk2} with $m_0 = 0.008\,$GeV, $m = 0.648\,$GeV and $\lambda = 0.9\,$GeV.  Dashed line: parameterization proposed in  Ref.~\cite{Oliveira:2020yac} of the LQCD calculations in Ref.~ \cite{Oliveira:2018lln}. 
Lower panel. The two quantities $ S^{V} (p^2)~[p^2-{\cal M}^2(0)]$  (solid line) and $ S^{S} (p^2)~[p^2-{\cal M}^2(0)]/{\cal M}(0)$ (dashed line)  vs. $p_E/{\cal M}(0)$.
} 
\label{fig:latheavy}
\end{figure}

{\it Dressing the quark propagator.}
As is well-known, in a given gauge, the dressed quark propagator can be written as follows 
\begin{equation}
    S(p)
    = i \, Z(p^2)~{\psla p +{\cal M}(p^2)\over p^2-{\cal M}^2(p^2)} ~,
\label{eq:SMZ}
\end{equation}
where ${\cal M}(p^2)$ is the quark-mass function and $Z(p^2)$  the wave-function renormalization. In the present work, disregarding the gauge dependence, we adopt  a phenomenological approach proposed in Ref. \cite{Mello:2017mor},  where $Z(p^2)=1$  and the quark-mass function ${\cal M}(p^2)$ was obtained  by fitting a recent LQCD quark mass-function \cite{Oliveira:2018lln,Oliveira:2020yac} \footnote{Notice that in Ref. \cite{Mello:2017mor}, the {\em Landau-gauge} LQCD  calculations of  Ref. \cite{Parappilly:2005ei}  were used.} by   using 
for $p^2 \le 0$ the following simple expression (with a pole in the timelike region)
\be
    {\cal M}(p^2)=m_0-{m^3\over p^2-\lambda^2+i\epsilon}~ ,
\label{eq:Mk2}
\ee
where the three parameters are chosen as follows: i) the bare mass $m_0$ and the infrared (IR) mass ${\cal M}(0)=m_0+ m^3/\lambda^2$ are taken equal to the values of  the accurate parameterization proposed in  Ref.~\cite{Oliveira:2020yac} of  the LQCD calculations presented in  Ref.~\cite{Oliveira:2018lln}, i.e. $m_0=m^{LQCD}_0=0.008$ GeV and ${\cal M}(0)={\cal M}^{LQCD}(0)=0.344$ GeV; ii)  while $\lambda$ is properly adjusted. In conclusion we have the following values in Eq. \eqref{eq:Mk2}
$$m_0=0.008\,GeV~, \quad m=0.648\,~GeV, \quad  \lambda=0.9\,GeV.$$
Notice that $m$ allows to interpolate between a current-mass quark  scenario and a fully-dressed one, while $\lambda$ yields the width at the half  height of  the difference ${\cal M}(0)-m_0$. The comparison between our fit and the mass function given in Ref.~\cite{Oliveira:2018lln} is shown in the upper panel of Fig.~\ref{fig:latheavy}, where in abscissa there is the Euclidean momentum $ p_E=\sqrt{-p^2}$.

The propagator in  Eq. \eqref{eq:SMZ} has poles whenever  $m_i={\cal M}(m^2_i)$.  In particular,  one has  only three poles, solutions of the cubic equation 
\begin{equation}
   m_i(m^2_i-\lambda^2)=\pm[ m_0(m_i^2-\lambda^2)-m^3] \, .
\label{eq:poles}\end{equation}
It turns out that $\rho^{V(S)}(s)$ in Eq. \eqref{eq:KLR} are  given  by a sum of three Dirac delta-functions \cite{Mello:2017mor}, viz.
\be
    \rho^{S(V)}(s) = \sum_{a=1}^{3} R^{S(V)}_a\delta(s-m_a^2)~, 
\label{eq:rho}
\ee
where
$R^{S(V)}_a$ are the residues,   that read  
\be
    R^{V}_a = \frac{(\lambda^2-m^2_a)^2}{(m_a^2-m_b^2)(m^2_a-m^2_c)}~, \nonu
    R^{S}_a = R^{V}_a~{\cal M}(m_a^2),
\label{eq:resid}\ee
with the indices $\{a,b,c\}$ following the cyclic permutation $\{1,2,3\}$.  They fulfill   the following relations
\be
    \sum_{a=1}^3R_a^{V}=1\,,\quad
     \sum_{a=1}^3R_a^{S}=m_0\, \,,
     \\ &&
     \sum_{a=1}^3m_a^2R_a^{V}=-2\lambda^2+\sum_{a=1}^3 m^2_a\,.
\ee
The actual values of both poles and residues are given in Table \ref{Tab:polres}.
\begin{table}[htb]
 \caption{Poles, $m_i$, and residues, $R_i$,(cf Eqs. \eqref{eq:rho} and \eqref{eq:resid}) for the fit to the LQCD mass function in Ref. \cite{Oliveira:2018lln}.  The IR mass ${\cal M}(0)=m_0+m^3/\lambda^2$ is 0.344\,GeV, and the parameters of the running mass are also given in the Table. }
  \label{Tab:polres}
\begin{center}
 \begin{tabular}{c c c c}
 \hline  \hline
  ~ $i$ ~ & $m_i/{\cal M}(0)$ & $R^V_i$ & $R^S_i/{\cal M}(0)$ \\ 
\hline
1  & 1.365 & 3.7784  & 5.1578\\
2  & 1.667  & -2.8863 & -4.8099\\
3  & 3.008  & 0.1079 & -0.3244  \\
 \hline
\hline
&$m_0/{\cal M}(0)$ & $~~m/{\cal M}(0)$ & $~~\lambda/{\cal M}(0)$
\\
\hline
& 0.0233 & 1.883 &
2.616\\
\hline
 \end{tabular}
 \end{center}
 \end{table}
Let us recall that for a physical particle belonging to the S-matrix representation, the 
K\"allen-Lehmann  spectral densities satisfy the positivity constraints~\cite{itzykson2012quantum}:
\begin{equation}\label{posab}
\rho^V(s)\geq 0 \,\, \,\,\text{and}\,\,\,\,\sqrt{s}\,\rho^V(s)-\rho^S(s)\geq 0 \, .
\end{equation}
Differently, in QCD the colored quark  cannot be  an asymptotic state and therefore  the weights in Eq.~\eqref{eq:KLR} should violate the positivity constraints, which actually happens in the present model.  For example,  if $m_a<m_b<m_c$ one has $R^V_b<0$, which  trivially implies   the violation of the positivity constraint $\rho^V(s)\geq 0$ for $s=m_b^2$. 

 The lower panel of Fig.~\ref{fig:latheavy} shows
 the two quantities $ S^{V} (p^2)~[p^2-{\cal M}^2(0)]$ and $ S^{S} (p^2)~[p^2-{\cal M}^2(0)]/{\cal M}(0)$ (cf. Eq.~\eqref{eq:KLR}, first line), obtained by means of Eq.  \eqref{eq:rho}, with  poles and  residues given in Table \ref{Tab:polres}. It is worth noting that, for both functions, the tails are the expected ones in the limit $p_E\gg {\cal M}(0)$, i.e. the ones pertaining to a  massless quark propagator, $\psla p/p^2$.
 
 {\it Solving the  $0^-$ BSE with dressed quarks.} To solve  the  BSE for a pseudoscalar $q\bar q$ bound-system, with momentum-dependent dressed quark-propagators, in Minkowski space, we apply the same technique already adopted in Refs.~\cite{dePaula:2016oct,dePaula:2017ikc}, but for the same BSE with  massive free-quark propagators.  We use the NIR~\cite{nak63} of  the scalar function $\phi_i(k,P)$, that reads
 \begin{equation}
\hspace{-.3cm}\phi_i(k,P)=\hspace{-.1cm}\int_{-1}^1 \hspace{-0.2cm}dz'\int_{0}^\infty \hspace{-0.2cm}d\gamma' { g_{i}(\gamma',z') \over \left[{k}^2+z' P\cdot k -\gamma'-\kappa^2+i\epsilon\right]^3},
\label{eq:NIR}
\end{equation}
where $\kappa^2 = m_1^2-M^2/4$, with $m_1$  the lightest pole (cf. Table \ref{Tab:polres}), and $g_{i}(\gamma',z')$ are the NWFs. The role of the NIR is to give to the BS amplitude an analytical structure in terms of the external momenta,  which makes it possible to analytically perform the needed loop-momentum integration present in the interaction kernel. Hence, from the initial BSE, through the so-called LF projection, i.e. the integration on the component $k^-=k^0-k^3$ (see details in  Refs.~\cite{dePaula:2016oct,dePaula:2017ikc}) and  taking into account Eq.~\eqref{eq:KLR}, one can formally obtain  the following system of integral equations for the NWFs  
\begin{multline}
\int_{0}^{\infty}d\gamma'\frac{g_{i}(\gamma',z)}{\Big[\gamma+z^2M^2/4+\gamma'+\kappa^2-i\epsilon\Big]^2}=
  \\ 
{\alpha \over 2\pi} \sum_{j}\int_{-1}^{1}dz'\int_{0}^{\infty}d\gamma' {\cal L}_{ij} (\gamma, z; \gamma', z') \, g_j(\gamma', z') \, .
\label{eq:BSEgij}
\end{multline}
where  $\alpha= {g^2/ 4\pi}$, with $g$   the dimensionless coupling constant,  $\gamma=|{\bf k}_\perp|^2$, and the kernel ${\cal L}_{ij}$ is obtained by a lengthy, but straightforward analytical steps (see Ref. \cite{Castro:2023} for details).

 It should be pointed out that if the system in \eqref{eq:BSEgij} admits a solution (N.B. one has to deal with a generalized eigenvalue problem) the use of the NIR is validated, and one can reconstruct the BS amplitude via Eq. \eqref{eq:NIR}.

{\it Results.}
 The solutions of the BSE for a  pseudoscalar $q\bar q$ bound-state  have been obtained for    masses  ranging from   $M= 0.653\,$GeV to $M=0.447\,$GeV, and suitable  values of both gluon mass $\mu$ and  vertex parameter $\Lambda$ to explore the interplay between these gluonic scales. The effective gluon mass is taken around $\sim 2\Lambda_{QCD}$~\cite{DuPRD14}, and the size of the quark-
gluon vertex of the order $\sim \Lambda_{QCD}$ (see, e.g., Ref.~\cite{Oliveira:2020yac}).

In previous studies  of the pion with  physical mass~\cite{dePaula:2020qna,dePaula:2022pcb}, by using  a fixed
quark mass of $0.255$\,GeV  (corresponding almost to the inflection point of the mass function shown in the upper panel of Fig.~\ref{fig:latheavy}), a gluon mass of $0.638$\,GeV and a
vertex parameter  $\Lambda=0.306\,$GeV, we obtained: i) a decay constant $f_\pi=130\,$MeV, in agreement with the experimental value~\cite{PDG_2018}  and ii) a  valence probability of 70\%, with 57\% for the antialigned combination of the quark spins and 13\% for the aligned  one (see below). In this case, the gluon contribution to the pion state from  the higher light-front Fock-components amounts  to 30\%, which reflects the fluctuation of the pion in a $q\bar q$ pair plus any number of gluons  (recall that we are using a ladder kernel).
Such massive gluon fluctuations  are generated by  a color source, which has  a size of about $\hbar c/ \Lambda = 0.645\,$fm 
(not far from the size of the pion) and is able to reproduce   the pion properties like the electromagnetic form factor~\cite{Ydrefors:2021dwa} and longitudinal parton distribution~\cite{dePaula:2022pcb}. In the present study we have a  IR mass of $0.344\,$ GeV   equal to the one of the  LQCD quark  mass-function~\cite{Oliveira:2018lln}, and we tune  the gluon mass to $\mu=0.469\,$ GeV and  $\Lambda=0.1\,$GeV, so that  the size of  the color source   considerably swells to $\sim \hbar c/ \Lambda \approx 2\,$fm. Thus, the gluon fluctuations are produced at large distances. Both the combined effect of the somewhat smaller gluon mass and the strong IR enhancement of the interaction  is expected to increase the Fock components of the pion containing the dynamical gluons and simultaneously   reduce the   valence probability. Indeed, at the pion mass, the fixed-mass approach, with a quark mass equal to $0.344$ GeV  and the previous values of $\mu$ and $\Lambda$, provides a percentage of 51\% for the Fock components beyond the valence one and a decay constant $f_\pi=89.8\,$ MeV. Due to  a quark-mass larger than the one in the previous studies,  the pion becomes more dilute, and the relevance of Fock components beyond the valence one increases from $30\%$  to $51\%$. 
Moreover,   the valence fraction of the aligned component reduces to 7\%, compared to the previous case of  18.6\% (stemming from the probability ratio $13/70$). 

To  gain a first insight into the dynamics generated  by a large color source when the running quark mass is considered,  with the corresponding parametrization  that favors a large gluonic content of the  bound state, i.e.  with $\mu=0.469\,$GeV and $\Lambda=0.1\,$GeV,  we have calculated the valence wave function, i.e.   the amplitude of the leading component in the Fock expansion of the state (see, e.g., Ref.~\cite{dePaula:2020qna})  for masses of the bound system in the range $3m_\pi<M<5m_\pi$ (cf. Table \ref{Tab:polVP}). The valence wave function  is obtained  by LF-projecting the BS amplitude, and  notably, it  can be decomposed into its quark-spin configurations, $S=0$ and $S=1$. One writes the  anti-aligned and aligned configurations, respectively, as \cite{dePaula:2020qna}
 \begin{equation}
 \begin{aligned}
\Psi_{\uparrow\downarrow}(\gamma,z)=&
\Psi_{\downarrow\uparrow}(\gamma,z)=
\psi_2(\gamma,z) +{z\over 2} \psi_3(\gamma,z)
\\
+ {i\over M^3} 
~\int_0^{\infty}& d\gamma'~
~{\partial g_{3}(\gamma',z;\kappa^2)/\partial z\over
  \gamma+\gamma'+z^2m^2 +(1-z^2)\kappa^2 } \,,
\\ 
\Psi_{\uparrow\uparrow} (\gamma,z)=& \Psi_{\downarrow\downarrow}(\gamma,z)=\frac{\sqrt{\gamma}}{M~} \psi_4(\gamma,z) \, , 
 \end{aligned}
\label{eq:psipar}
\end{equation}
where  $m=m_1$ for the running mass case, and the amplitudes  $\psi_i(\gamma,z)$ are obtained by integrating on $k^-$ the scalar functions in Eq.~\eqref{eq:NIR}. In correspondence, one defines
the valence probability density as follows
\begin{equation}
\rho_{val}(\gamma,z)=\rho_{\uparrow\downarrow}(\gamma,z)
+\rho_{\uparrow\uparrow}(\gamma,z) \, ,
\label{eq:rhoval}
\end{equation}
where
\begin{eqnarray}
&&\rho_{\uparrow\downarrow(\uparrow\uparrow)}(\gamma,z)=\frac{N_c}{16\pi^2}
   |{\Psi}_{\uparrow\downarrow(\uparrow\uparrow)}(\gamma,z)|^2\,  \, .
 \label{eq:pval}
\end{eqnarray}
 Finally, by normalizing   $\rho_{val}$ to $1$,  we have calculated: i) the percentage of each spin components, $P_{\uparrow\downarrow}$ and $P_{\uparrow\uparrow}$, and  ii) the  longitudinal and transverse distributions, that in turn can be decomposed in their spin components. Namely,
the longitudinal and the transverse momentum distributions are given by
\be
\phi(\xi)= \phi_{\uparrow\downarrow}(\xi)+\phi_{\uparrow\uparrow}(\xi)\, ,
\label{eq:longdis}
\\ &&
{\cal P}(\gamma)= {\cal P}_{\uparrow\downarrow}(\gamma)+{\cal P}_{\uparrow\uparrow}(\gamma)\,,
\label{eq:trandis}
\ee
where 
\be 
\phi_{\uparrow\downarrow(\uparrow\uparrow)}(\xi)=\int^\infty_{0}d\gamma ~
\rho_{\uparrow\downarrow(\uparrow\uparrow)}(\gamma,z)\, , 
 \label{eq:longspin}
 \\ &&
 {\cal P}_{\uparrow\downarrow(\uparrow\uparrow)}(\gamma)=\int^1_{-1} dz~
\rho_{\uparrow\downarrow(\uparrow\uparrow)}(\gamma,z)\, .
\label{eq:transpin}
\ee
with $\xi=(1-z)/2$.

\begin{table}[h]
\caption{The  chosen masses, $M$,  of the   $0^-$ bound-system (in unit of the IR mass ${\cal M}(0)=0.344$ GeV)   are presented  along with the  coupling constants $\alpha=g^2/4\pi$ and the percentages of the spin configurations in the valence wave function.  Recall that in addition to $M$ the set of   model parameters is completed by : i) the gluon mass $\mu/\mathcal{M}(0)=1.363$, ii)  a the vertex parameter $\Lambda/\mathcal{M}(0)=0.291$ and iii) $\lambda/\mathcal{M}(0)=2.616$ (see  Eq. \eqref{eq:Mk2}).  For each $M$, the first line represents the dressed case, while the second line is the undressed one with a quark mass equal to $0.344\,$GeV.}
 
  \label{Tab:polVP}
\begin{center}
 \begin{tabular}{c c c c c }
 \hline  \hline
${M}/{{\cal M}(0)} $  & $g^2$ & $\alpha$ & $P_{\uparrow\downarrow} (\%) $  &  $P_{\uparrow\uparrow} (\%)$  \\
\hline
1.9&7.62&0.61&93&7\\
~  &3.76&0.30&96&4\\
\hline
1.6&12.46&0.99&93&7\\
~  &11.29&0.90&93&7\\
\hline
1.5&14.13&1.12&93&7\\
~  &13.67&1.09&93&7\\
\hline
1.4&15.78&1.26&94&6\\
~  &15.93&1.27&93&7\\
\hline
1.3&17.38&1.38&94&6\\
~  &18.07&1.44&93&7\\
\hline
1.2&18.93&1.51&94&6\\
~  &20.06&1.60&93&7\\
 \hline
 \end{tabular}
 \end{center}
 \end{table}

Table~\ref{Tab:polVP},  for given values of $M$, $\mu$ and $\Lambda$ shows some  outputs of our calculations:     coupling constant and  percentages of the spin configurations in the valence wave function.  In general, the pattern  for  running and fixed mass cases is the expected one: when the mass of the system decreases the coupling constant increases. Loosely speaking, if the binding energy, i.e. the difference between the bound-system mass, $M$, and some typical constituent mass   increases  also the depth of the well does.  Heuristically, by defining the binding energy as $B=2m_q-M$ where $m_q$ is an effective quark mass,  one can choose $m_q=\mathcal  {M}(0)$, for the fixed mass case,  while one necessarily has $m_q=\bar m \le \mathcal {M}(0)$, for the running mass one. In the latter case, the effective mass  $\bar m $ stems from the suitable combination of the mass function, shown in the upper panel of Fig.~\ref{fig:latheavy}, and the quark momentum-distribution, dictated by the BSE. Hence,  one expects a decreasing $\bar m$   when  the binding increases, since   the system shrinks and the momentum-distribution tail grows, emphasizing the small mass region in ${\cal M}(p^2)$. In conclusion for given $M$,  the coupling constant for the running-mass case is larger than the one for the fixed-mass case  and increases more slowly
($\bar m$ saturates to some value in the IR region).

\begin{figure*}[t]  
\includegraphics[width=6cm]{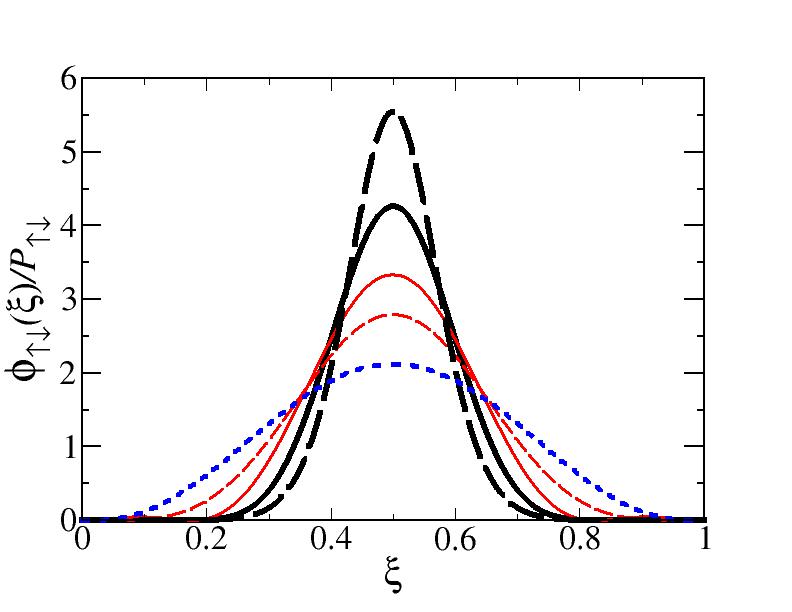}
\hspace{-.5cm}
\includegraphics[width=6cm]{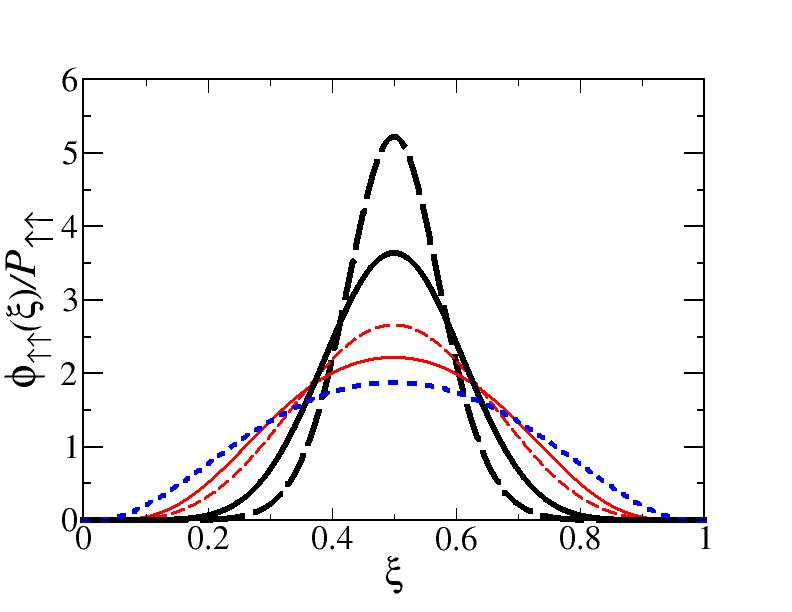} 
\hspace{-.5cm}
\includegraphics[width=6cm]{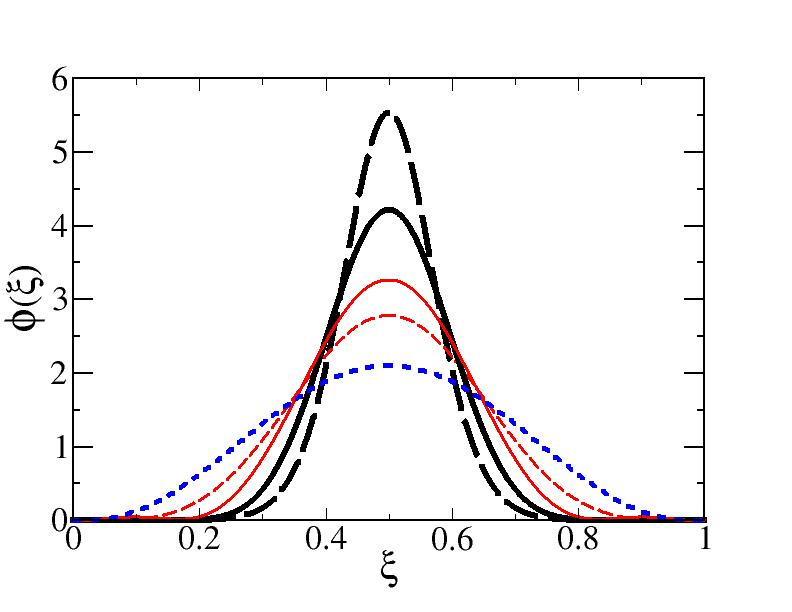}
    \caption{Longitudinal momentum distributions defined in Eqs.~\eqref{eq:longdis} and \eqref{eq:longspin}, with  $\Lambda = 0.1\,$GeV and $\mu = 0.469\,$GeV. Thick solid line: running mass model for  $M=0.653\,$GeV. Thin solid line: the same as the thick one, but for  $M=0.447\,$GeV.
    Thick dashed Line: fixed quark mass equal to  $0.344\,$GeV and  $M=0.653\,$GeV.  Thin dashed line: the same as the thick one, but for $M=0.447\,$GeV. Dotted line: the same as the thick one, but for $M=m_\pi=0.141\,$GeV.}
    \label{fig:Long}
\end{figure*}

\begin{figure*}[t]  
\includegraphics[width=6.cm]{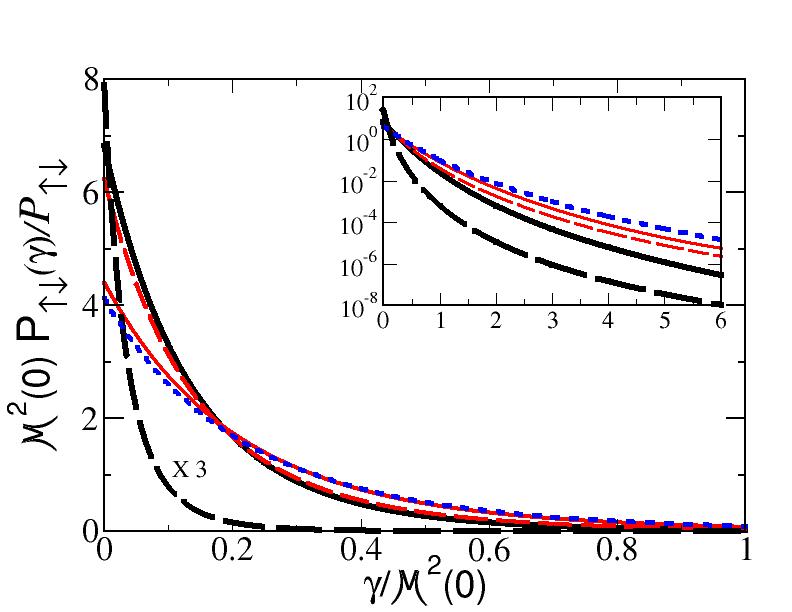}
\hspace{-.4cm}
\includegraphics[width=6.cm]{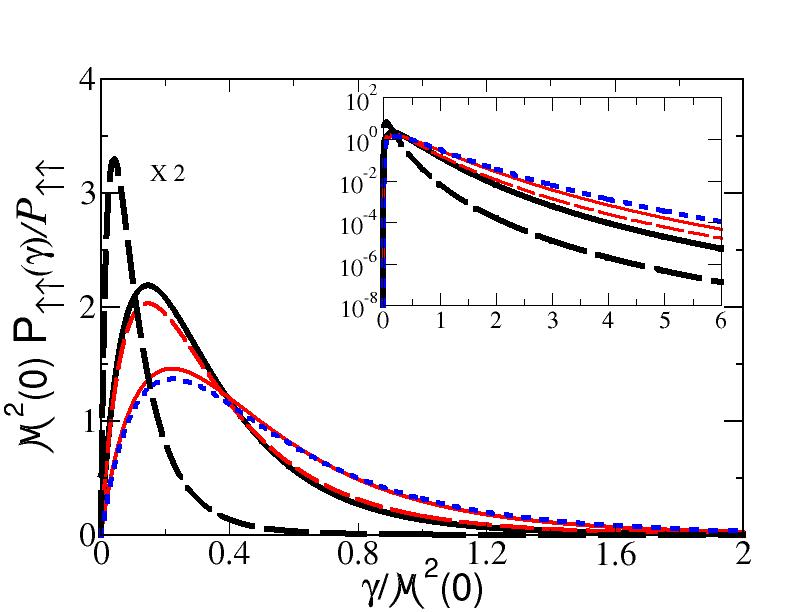} 
\hspace{-.4cm}
\includegraphics[width=6.cm]{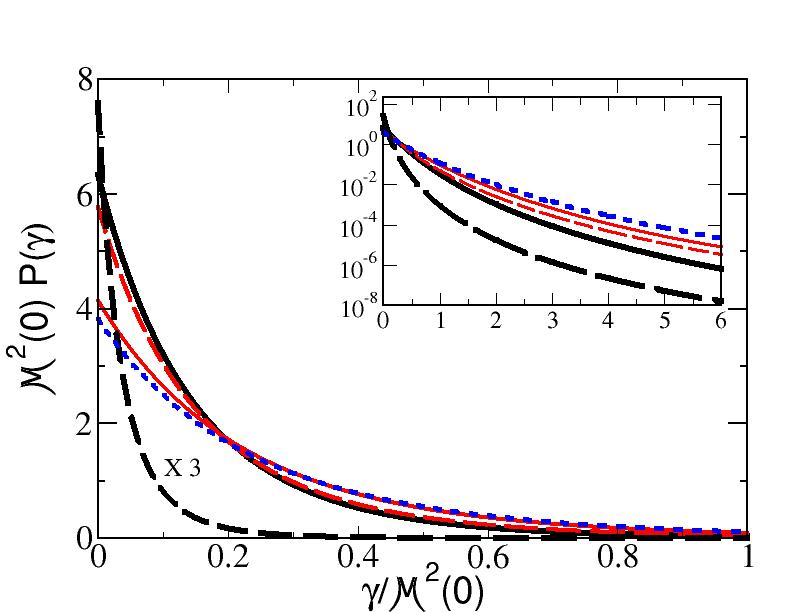}
    \caption{Transverse momentum distributions defined in Eqs. \eqref{eq:trandis} and \eqref{eq:transpin}, with  { $\Lambda = 0.1\,$GeV and  $\mu = 0.469\,$GeV.} The legend of the lines is the same as in Fig.~\ref{fig:Long}.}
    \label{fig:transv}
\end{figure*}

Table~\ref{Tab:polVP}  also shows  the relative weight in the valence state between the  anti-aligned spin component, which is dominant (recall that one has an orbital-angular momentum equal to $0$ for a spin singlet), and the aligned spin component. 
On the spanned range of the mass $M$, the two spin configurations keep almost the same percentage for both running and fixed quark masses. This could be interpreted as a feature dictated by the adopted interaction kernel.

In Figs.~\ref{fig:Long}
and \ref{fig:transv}, the longitudinal and transverse distributions for the two examples of $M/{\cal M}(0)=1.9$ and $1.3$ are shown for the momentum-dependent dressed case (solid lines) and the  undressed one (dashed lines). 

As shown in Fig.~\ref{fig:Long}, for the largest value of the system mass, $M/{\cal M}(0)=1.9$, the   total longitudinal distribution  for the fixed-mass case is narrower than the running-mass one, implying that it  decreases faster when approaching the end-points. 
Such a property 
can be understood by smaller value of the coupling constant (cf. the first two lines in Table  \ref{Tab:polVP}) that leads to a larger size of the system and hence a smaller average relative momentum. Independently of the dressing of the quarks and bound state mass, the aligned longitudinal momentum distribution is broader than the anti-aligned one due to the relativistic nature of the former.  For $M/{\cal M}(0)=1.3$, the running mass-case and the fixed-mass one yield similar results, as one could expect given the similar values of the corresponding coupling constants (cf. Table  \ref{Tab:polVP}).

The transverse distributions in Fig.~\ref{fig:transv} show that the dressing of the quark-propagator generates a larger momentum tail than in the undressed case, regardless of  the spin component,  but it decreases when the mass-system does.  Moreover, the effect of  the dressed quark-gluon vertex  is manifested in the  damping of the kernel at a scale of $\Lambda^2/\mathcal{M}^2(0)\sim\gamma/\mathcal{M}^2(0) \sim 0.1$,   as clearly recognized in the plots for the larger binding corresponding to $M/\mathcal{M}(0)=1.3$ in both  dressed and undressed quark cases. Finally, one should notice that the aligned transverse distribution vanishes for $\gamma=0$, as expected for an orbital-angular momentum $L=1$ component.
The insets in Fig.~\ref{fig:transv} show the tail of the transverse distributions which acquire a close functional dependence on  $\gamma$, as it is expected, once the dressed and undressed propagators for large momentum are dominated by their gluon component and tend to be identical (cf. bottom panel of Fig.~\ref{fig:latheavy}).  

{\it Summary.}
We  have solved the $0^-$ homogeneous Bethe-Salpeter equation  in Minkowski space by using i) the NIR of the BS amplitude; ii) an extended interaction kernel based on the ladder exchange of  massive gluons, in the Feynman gauge, and iii) a phenomenologically dressed quark propagator, already adopted in Ref.~\cite{Mello:2017mor}. The quark-mass function has only one adjusted parameter that  was  tuned on LQCD calculations of the running mass in  Ref.~\cite{Oliveira:2018ukh}. It has been chosen a gluon mass  close to $2\Lambda_{QCD}$ and an extended color density with size $\sim 2\,$fm, that enforces the gluonic content of the bound system. 
The explored range of the system mass   was  $3m_\pi<M< 5m_\pi$, and we obtained both static (cf. Table \ref{Tab:polVP}) and dynamical (cf. figs. \ref{fig:Long}  and \ref{fig:transv}) quantities. The main features have been discussed and heuristically interpreted, in order to enhance our physical intuition of the outcomes of our approach, that soon  will be extended to the realistic case of the pion. In particular, we stressed the role of the interplay between three gluonic phenomena: dressing of the quark propagator, extension of the quark-gluon vertex and ladder exchange of massive gluons. For instance, we found that the quark dressing widens the transverse momentum distribution when compared to the case of the undressed quark tuned to have the IR mass, and 
interestingly, the aligned spin component of the valence wave function is suppressed with respect to the anti-aligned one, regardless of whether a  running  quark-mass is used or not, pointing to a  ladder kernel effect.

Plainly, the present   application  of our approach based on the BSE plus a phenomenological running quark-mass should be considered as an initial step towards  a description of the strongly-bound pion within a more consistent framework, based on realistic solutions of the quark gap-equation able to yield ${\cal M}(p^2)$ directly in Minkowski space (see also, e.g., Refs.~\cite{Bicudo:2003fd,Sauli:2006ba,Siringo:2016jrc,Tanaka:2017sdd,Biernat:2018khd,Solis:2019fzm, Mezrag:2020iuo,Duarte:2022yur}).

{\it A. C. gratefully thank
INFN Sezione di Roma
 for providing the computer resources to perform all the calculations shown in this work. W. d. P. acknowledges the partial support of CNPQ under Grants No. 313030/2021-9 and the partial support of CAPES under Grant No. 88881.309870/2018-01. T. F. thanks
the financial support from the Brazilian Institutions:
CNPq (Grant No. 308486/2015-3), CAPES (Finance Code 001) and FAPESP (Grants 
No. 2017/05660-0 and 2019/07767-1). E.
Y. acknowledges the support of FAPESP Grants No. 2016/25143-7 and No. 2018/21758-2. This work is a part of the
project Instituto Nacional de  Ci\^{e}ncia e Tecnologia - F\'{\i}sica
Nuclear e Aplica\c{c}\~{o}es  Proc. No. 464898/2014-5.}

\bibliography{piontmd}

\end{document}